# IN-SITU X-RAY PHOTOEMISSION STUDY OF NB SURFACE UNDER PLASMA CLEANING DURING MID-T BAKING FOR SRF CAVITIES*

C. Boutelaa[1,2]†, S. Gruszka[1], C. Cheney[1], J. Yemane[1], T. Gerardin[1], E. Mistretta[1], J. Demailly[1], R. Laxdal[2], P. Kolb[2], J. Keir[2], B. Mercier[1], N. Prud'homme[3], D. Longuevergne[1], G. Sattonnay[1]

[1]IJCLab, Paris Saclay University, Orsay, France
[2]TRIUMF, Vancouver, Canada
[3]ICMMO, Paris Saclay University, Orsay, France

*Abstract*

Specific heat treatments applied to superconducting radio-frequency (SRF) cavities, such as nitrogen infusion or Mid-T baking, aim to improve the quality factor ($Q_o$) at medium accelerating fields (˜10–20 MV/m). These treatments reduce the BCS surface resistance by tuning the mean free path of niobium over a few hundred nanometers, either by diffusing oxygen from the native oxide layer or by diffusing nitrogen after the dissolution of the oxide layer. However, these treatments preclude the usual chemical polishing, as it would reverse the beneficial effects of the heat treatments, making the cavities highly sensitive to surface contamination. In particular, the formation of niobium carbides, which can mask the expected benefits, strongly depends on the annealing conditions, surface preparation, and the material's history. To better understand these phenomena, niobium samples was annealed under ultrahigh vacuum (Mid-T baking) with $Ar/O_2$ plasma treatment to investigate surface contamination with *in-situ* heat treatment at 500°C and XPS analysis.

## INTRODUCTION

The preparation of cavity inner surfaces has been one of the major challenges in superconducting radio-frequency (SRF) accelerator technology. Various surface treatment methods have been proposed and explored to push to higher $Q_o$ at high accelerating gradient $E_{acc}$ of SRF cavities in recent decades. A high temperature annealing at 800°C followed by a medium temperature baking (mid-T bake) at 300°C during 3h after re-oxydation, is a novel surface doping procedure, enabling $Q_o$ values larger than $1\times10^{10}$ at 2.0 K but limiting the maximal achievable accelerating gradient at 25 MV/m [1,2]. The performance improvements associated with the Mid-T bake treatments have been attributed to the dissolution of the native oxide layer and the subsequent diffusion of oxygen into the near surface, which provides a positive doping effect [3]. It has also been observed that mid-T treatments are associated with carbide formation depending on the temperature of the treatment (see Fig. 1) [4].

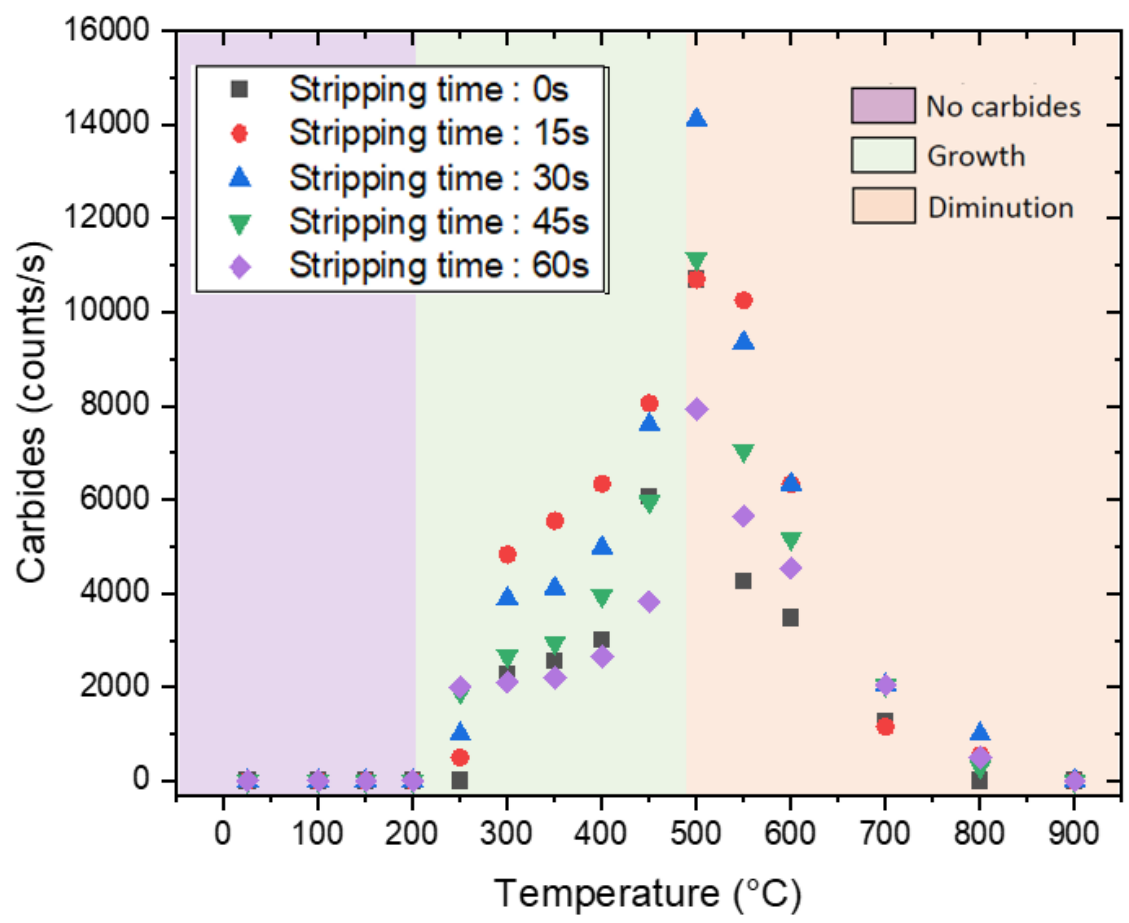

Figure 1: Evolution of niobium carbides from 25 °C to 900 °C as analyzed by XPS [4].

The carbon contamination, niobium carbides (NbC, $Nb_2C$) and oxide layer formation under $Ar/O_2$ plasma cleaning were analyzed after 500°C heat using in-situ X-ray photoelectron spectroscopy (XPS) and ex-situ scanning electron microscopy (SEM). We have chosen 500°C instead of the typical 300°C in order to amplify the phenomenon and simulate the worst-case scenario of the mid-T treatment determined in an earlier study [4].

## I. EXPERIMENT

### I.1 Setup

The *in-situ* Plasma-XPS-Baking experiments were performed in the Vacuum and Surfaces Platform at IJCLab. The set-up is equipped to perform *in-situ* XPS studies during annealing under UHV, with a dedicated plasma chamber as shown schematically in Fig. 2.

* Work supported by the CNRS-TRIUMF International Research Laboratory NPAT (Nuclear Physics, Nuclear Astrophysics and Accelerator Technology)

† chahinez.boutelaa@ijclab.in2p3.fr, https://orcid.org/0009-0009-2206-8579

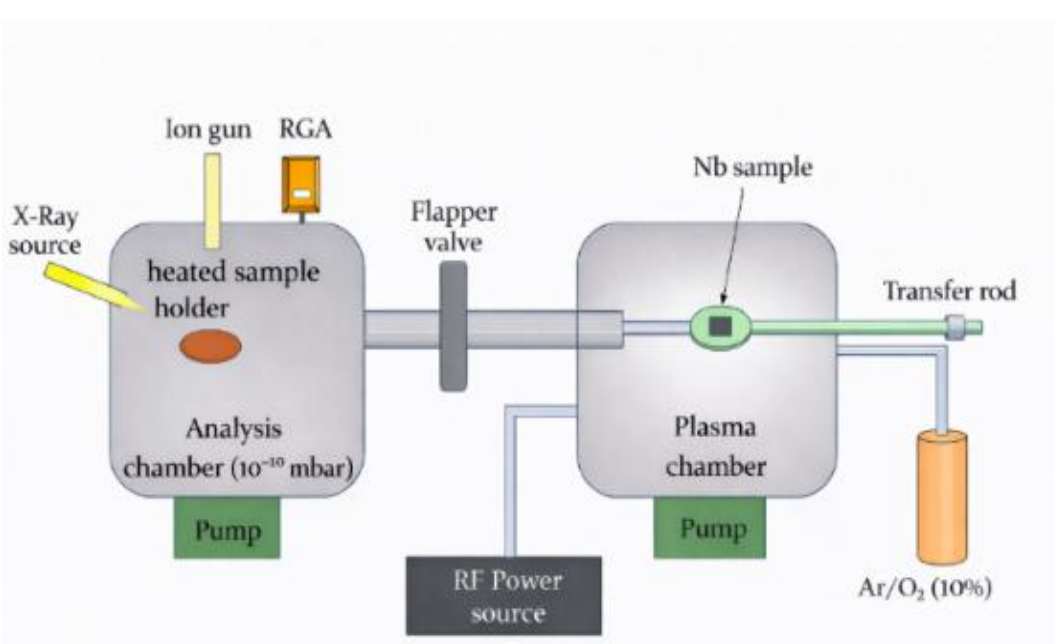

Figure 2: Experimental set up for plasma cleaning, XPS analysis, and sample annealing.

I.2 Nb preparation

Bulk niobium (Nb) used for SRF cavities consisted of fine-grain sheets with a residual resistance ratio (RRR) above 250. At TRIUMF, the material was cut into 10 × 10 mm samples and chemically polished using a BCP solution ($HF:HNO_3:H_3PO_4$ = 1:1:2), removing about 100 µm of surface material. The samples were then rinsed with ultra-pure water. Heat treatments were performed in the TRIUMF induction furnace at 800 °C for 3 hours under UHV ($10^{-7}$ Torr). Afterward, a flash BCP etching of about 10 µm and additional ultra-pure water rinsing were carried out, including a final rinse at IJCLab before the experiment. Two niobium samples followed different procedures (Fig. 3). In procedure A, the sample was first analyzed by XPS as a baseline measurement, then subjected to a mid-T bake corresponding to annealing at 500 °C for 3 hours at $10^{-9}$ mbar (Mid-T bake). The second experimental procedure (procedure B) was applied to the second sample. which consisted of the same steps but the sample was exposed to plasma treatment. The sample was cleaned with an argon and oxygen (10%) plasma mixture for 1h with power output corresponding to approximately 50 W (Plasma). Then, the sample was transferred to the analysis chamber for XPS measurement and heat treatment was perform for 500°C with the same condition that procedure A (Plasma + Mid-T bake).

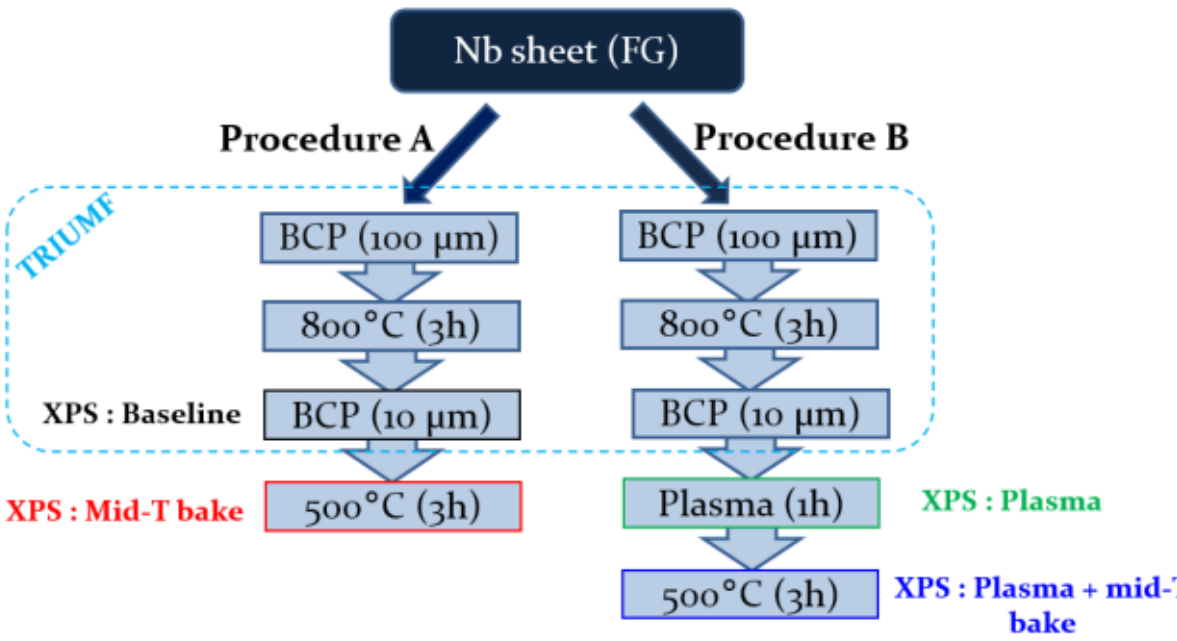

Figure 3: List of steps for the two different experimental procedures.

## II. RESULTS

II.1 Plasma impact on Nb carbide during mid-T baking

For the baseline, the C1s spectrum is characteristic of the typical adventitious carbon contamination commonly observed on metallic surfaces (Fig. 4). It exhibits the three main contributions generally associated with atmospheric exposure. The dominant component, centred at 285.0 eV, corresponds to C–C bonds, secondary components located at approximately 287.7 eV (C=O) and third at 289.0 eV (O–C=O) reveal the presence of oxygen-containing carbon species originating from atmospheric contamination. After annealing at 500 °C (Mid-T bake) of the procedure A, the C1s spectrum shows a component at 284.3 eV attributed to graphitic carbon and another component at 282.3 eV, which corresponds to Niobium carbide. Measurements after plasma processing in the procedure B clearly show an effective removal of carbon contamination, including the C–C, C=O, and O–C=O components. After the plasma treatment and the XPS measurement, the sample was directly annealed at 500 °C for 3 h. The C1s spectrum (Plasma + Mid-T bake) reveals the presence of a graphitic carbon at 284.2 eV [5], as well as a Niobium carbide at 281.9 eV.

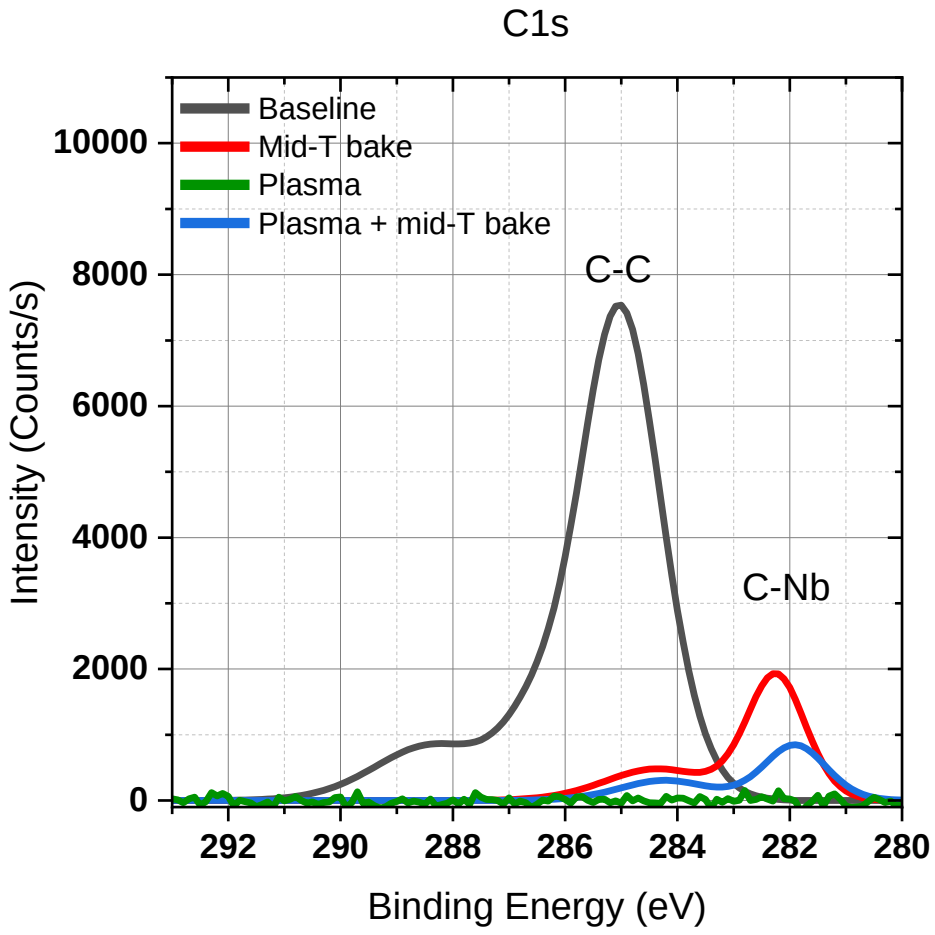

Figure 4: Comparison of C1s peaks for the both procedures.

The plasma treatment prior to annealing reduced the formation of niobium carbides, as evidenced by a decrease in the intensity of the carbide component, but also by a shift in binding energy of approximately 0.4 eV, as shown by the red and blue spectra in the figure 4. This shift may be associated with a phase transition of the carbide from $Nb_2C$ [6] to NbC [7]. These structures may also be observed by SEM.

II.2 Carbides structure by Scanning Electron Microscope (SEM)

In an effort to clearly identify the two phases, the samples have been analyzed by *ex-situ* SEM after air exposure. The surface morphology information for the Baseline (a), Mid-T bake (b), Plasma treatment followed by a baking (c) is represented in the figure 5.

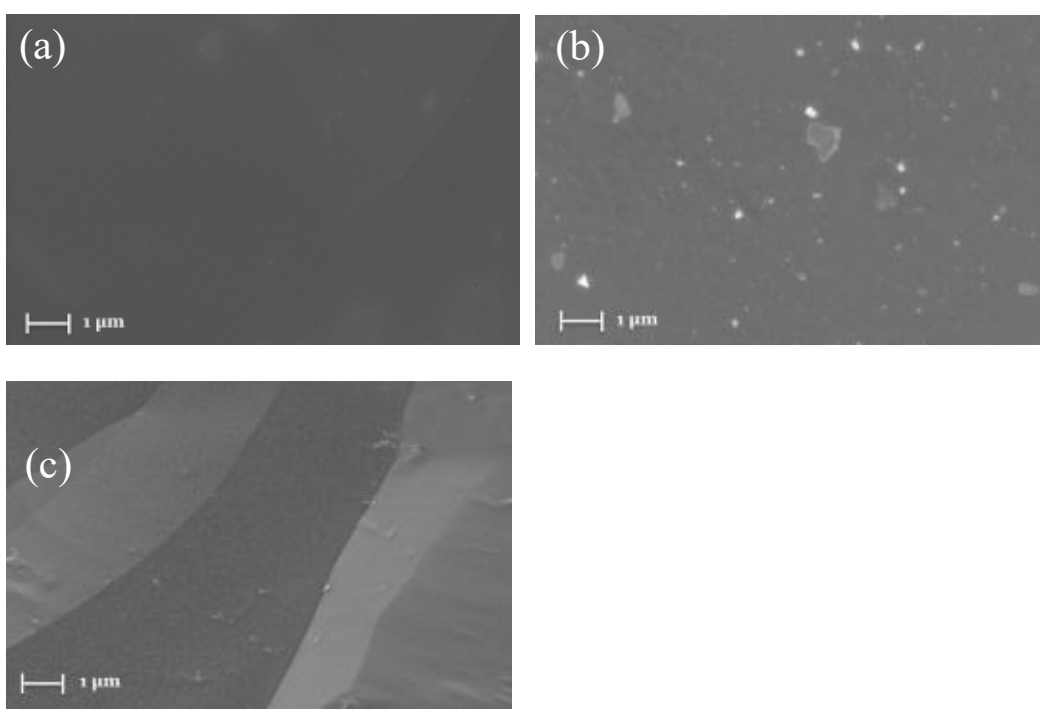

Figure 5: SEM pictures of baseline (a), Mid-T bake (b) and plasma + Mid-T bake (c).

The baseline, figure.5(a), free of carbides, shows the Niobium surface without any precipitates. The image in Fig. 5 (b) demonstrates the submicrometer-size pyramid-shape precipitates which are assigned to a $Nb_2C$ phase [8]. Fig. 5(c), fine white spots are observed, which are characteristic of the NbC phase. Previous studies [9], showed that the presence of the $Nb_2C$ phase is detrimental for superconducting properties inducing a steady Q-slope, sign of the degradation of the surface resistance versus accelerating gradient. On the other hand, NbC phase, obtained in low carbon contamination environment, because of its superconducting properties, may not be problematic, or may, on the contrary, participate to the improvement of $Q_o$ factor [4].

II.3 Plasma impact on Nb oxide during mid-T baking

The spectra of O1s for the both procedures were obtained in Fig. 6. The signal of the baseline recorded at 530.4 eV corresponds to $Nb_2O_5$ and is also visible in the Nb 3d spectrum at 207.0 eV for the 5/2 orbital and 209.7 eV for the 3/2 orbital in the figure 7. After Mid-T baking, the oxide is significantly dissolved but still visible in the O1s spectra with red peak. After plasma treatment, the O1s and Nb3d spectra show a clear energy shift, respectively from 530.4 eV to 529.5 eV and from 207.0 eV to 206.1 eV, corresponding to a reduction of the $Nb_2O_5$ into $NbO_2$ oxide [10]. This shift is also observed each time the sample was directly exposed to the plasma. This reduction of the native is rather counter-intuitive as the plasma is mainly composed of oxidizing elements. Moreover, the higher amplitude of $NbO_2$ peak compared to $Nb_2O_5$ peak is suggesting that the reduction of the oxide is linked to the thickening of the oxide layer and thus diffusion of oxygen into the bulk.

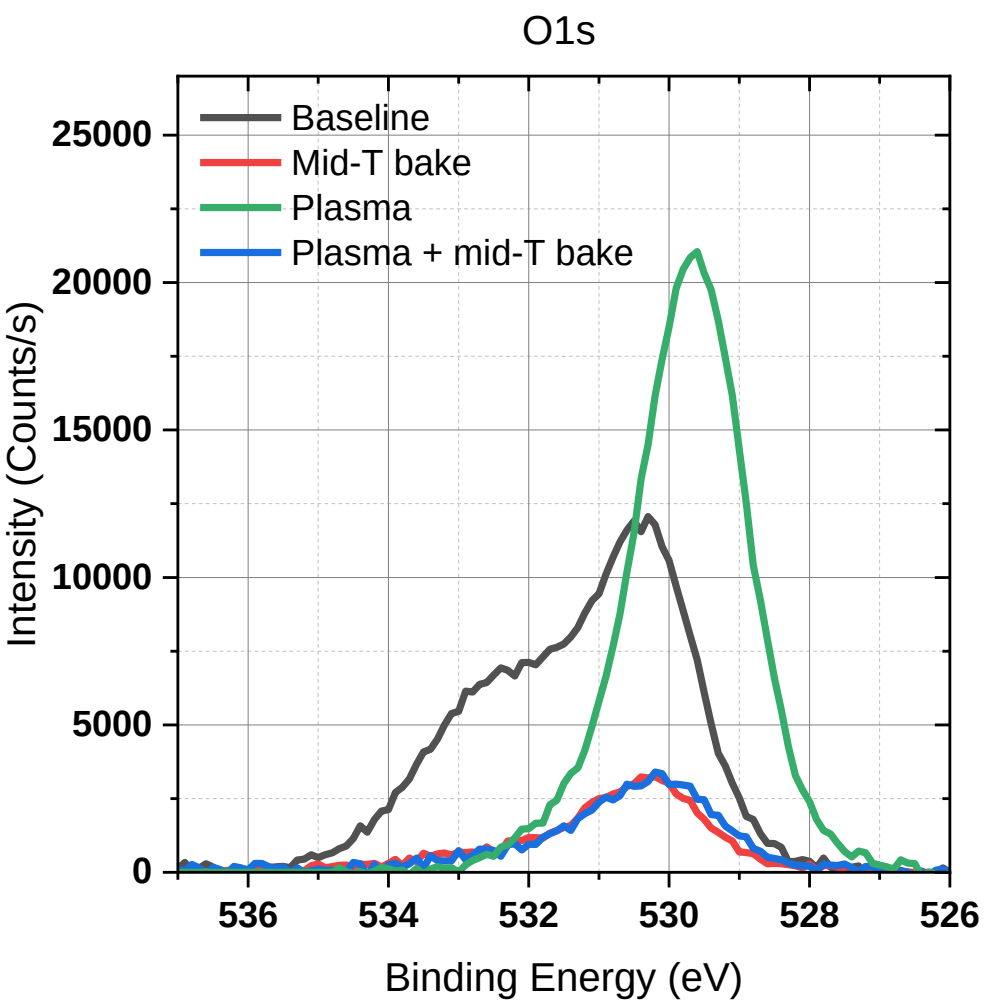

Figure 6: Comparison of O1s peaks for the both procedures.

A clear difference in doublet peak shapes in the Fig. 7 is visible meaning that the plasma treatment prior to heat treatment has non negligeable impact on the surface stochiometry of Niobium. For sample following procedure A, the Nb 3d XPS spectrum reveals a NbO phase at 202.8 eV for the 5/2 orbital, $Nb_xO$ ($1 < x < 2$) at 202.4 eV, metallic niobium (Nb°) at 201.7 eV [10] and $Nb_2C$ at 204.0 eV. For the sample following procedure B, when plasma treatment is applied prior to annealing (Blue peak), the surface becomes significantly more metallic. Nb 3d XPS spectrum reveals a NbO phase at 202.8 eV for the 5/2 orbital, metallic niobium (Nb°) at 201.7 eV and NbC 203.8 eV.

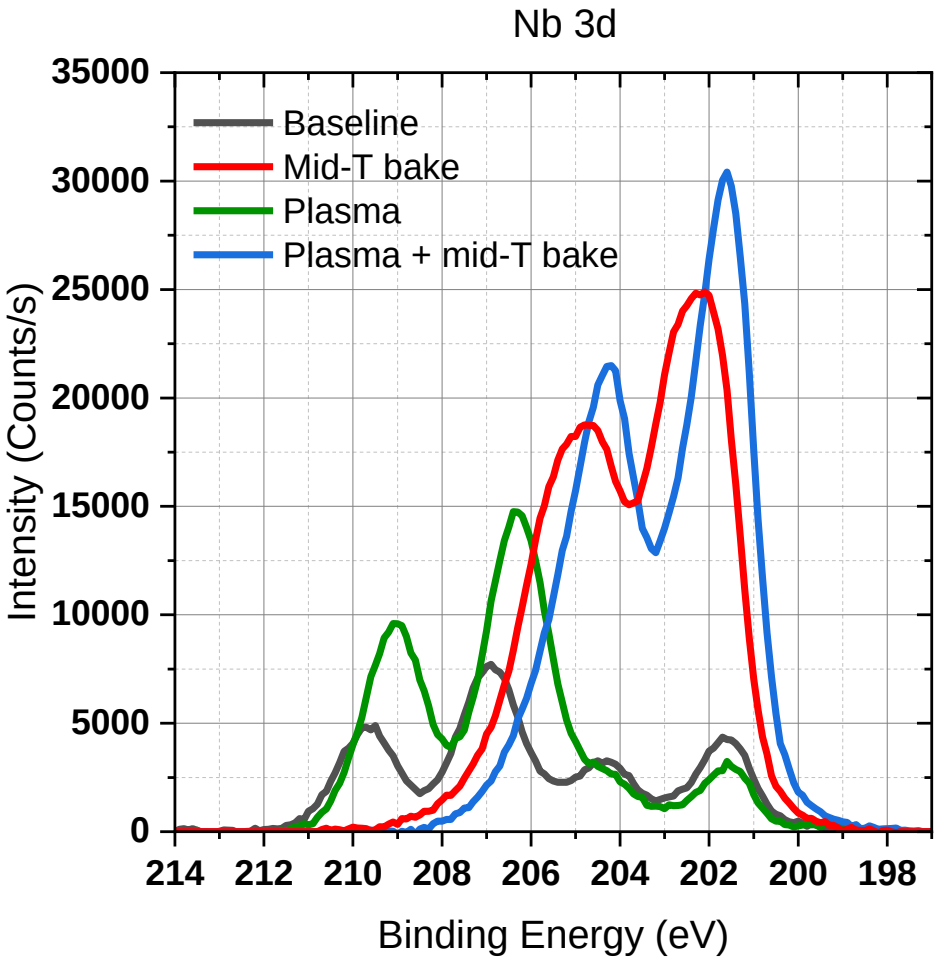

Figure 7: Comparison of Nb3d peaks for the both procedures.

## CONCLUSION

This study showed that the removal of carbon contamination and the thickening of the oxide layer prior to heat treatment thanks to a reactive plasma treatment is promoting the formation of NbC carbides instead of $Nb_2C$. Plasma treatment before heat treatment allowed the elimination of $Nb_xO$ sub-oxide and the increase of metal proportion. This preventive treatment may also improve considerably the surface quality of Niobium when used as a substrate before thin film deposition or for Q-bit applications.

## ACKNOWLEDGMENT

This work was supported by CNRS Nuclei&Particles Institute and the CNRS-TRIUMF International Research Laboratory NPAT (Nuclear Physics, Nuclear Astrophysics and Accelerator Technology). This work was as well supported by state aid administered by the National Research Agency (ANR) under the Investments for the Future Program, reference number ANR-21-ESRE-0049. The authors would like to thank the members of the platform "Vide&Surfaces" at IJCLab and of the SRF team at TRIUMF for the technical support and the preparation of Nb samples.